\documentclass[doublecol]{epl2} 
\usepackage{graphicx}

\title{Anomalous Relativistic Tunneling  and Exotic Point Interactions}

\author{
Pavel Hej{\v c}{\'i}k
\and 
Taksu Cheon
}

\institute{                    
Laboratory of Physics,
Kochi University of Technology,
Tosa Yamada, Kochi 782-8502, Japan
}

\abstract{
We examine one-dimensional quantum scattering of a Dirac particle off relativistic potential barriers.
With proper considerations of Dirac sea, existence of anomalous tunneling at zero incident-energy 
is revealed for a particular type of relativistic potential having same magnitudes and opposite signs 
for scalar and vector components. 
It is also shown that this leads to an exotic short range limit of the potentials.
}

\begin{document}

\maketitle

Since its conception, Dirac equation has been plagued by difficulties related to the {\it Dirac sea},
and it is only after the proper taming of
this monster of negative continuous spectra, that useful results are extracted from it.
This fact is demonstrated again recently by two example, one 
on the resolution of Klein paradox \cite{KL29,KS04}, and the other  
on the resolution of 
Plesset's no-bound state problem  \cite{PL32,TI61,GS07}.
In light of the resurgent interest in the Dirac equation in condensed matter physics \cite{KV94, KN07},
as well as of traditional interest in nuclear physics \cite{SW86}, we believe it is timely to report, 
in this Letter, a new result concerning a subtle effect of negative spectra on simple
one dimensional potential scattering of Dirac particle off relativistic potential barriers.

Here, we show that proper consideration of Dirac sea has
intriguing ramifications for low energy scattering matrices.  
In particular, it is shown to cause anomalous tunneling at zero incident energy 
for  ``$S=-V$ potentials'',
a particular type of relativistic potential having the same magnitudes and opposite 
signs for scalar and vector components of the potential. 

We further show, by considering the short-range limit of  this potential scattering,
that it leads to exotic low-pass gaussian wave filter 
whose non-relativistic kinematics limit is ``delta-prime'' point interaction, that
causes discontinuity in quantum wave function itself  but not in its derivative \cite{SE86},
which is distinct from conventional delta-function point interaction.


We start by considering Dirac equation in one dimension
that takes the following two-component form
\begin{eqnarray}
\label{e05}
\pmatrix{ \varphi' \cr \chi' }
=
\pmatrix{ 0 & m+\varepsilon+S-V\cr m-\varepsilon+S+V & 0 } 
\pmatrix{ \varphi \cr \chi }.
\end{eqnarray}
where $m$, $\varepsilon$ stand for the mass and relativistic energy of a Dirac particle,
and $S$ and $V$ are scalar and (time component of) vector potentials.  We only treat
time-symmetric systems, so the spatial components of vector potential is absent.
The {\it prime} signifies the spatial derivative $\frac{d}{dx}$.
%
%
%
\begin{figure}
\center{ \includegraphics[width=6cm]{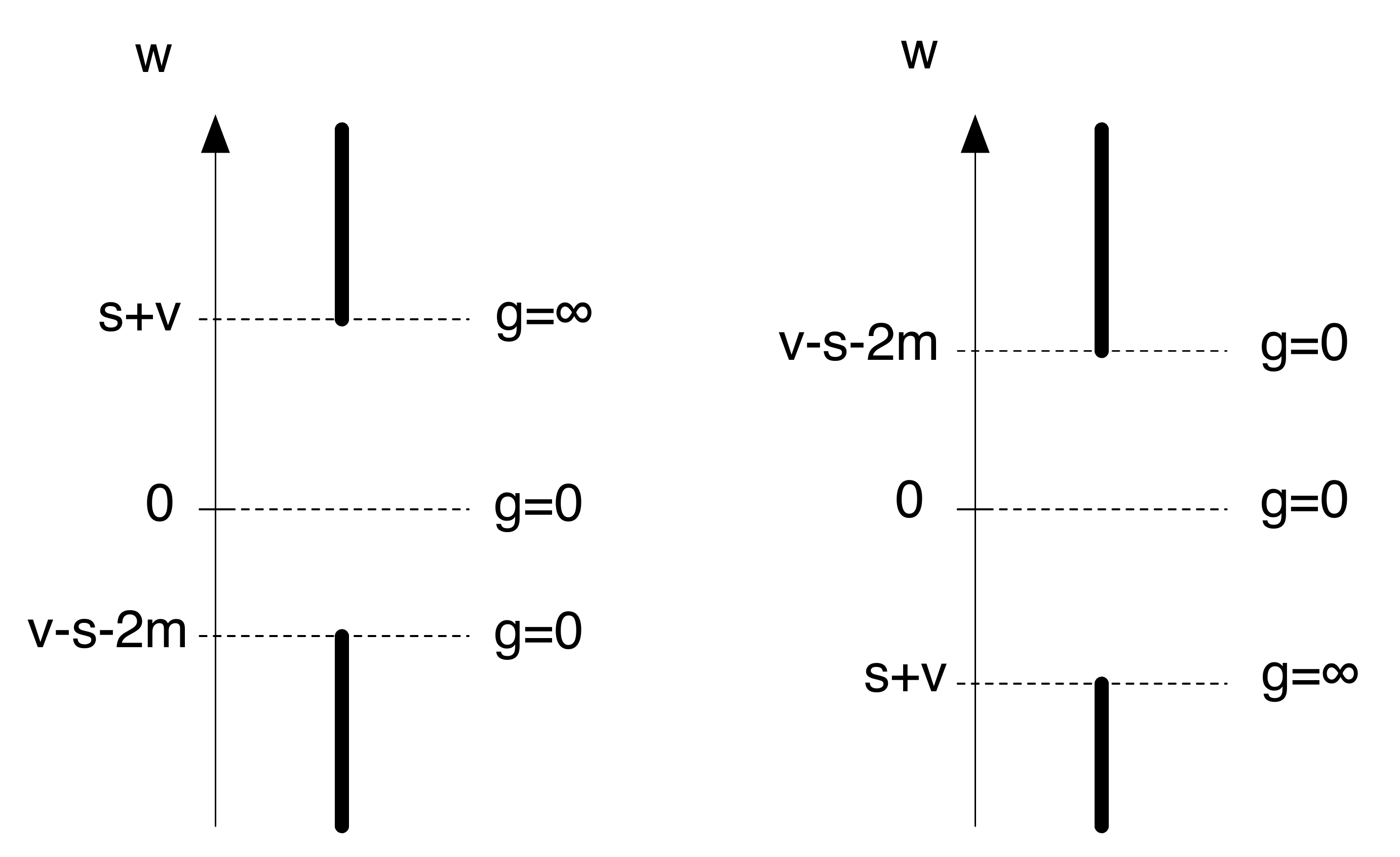} 
 }
\label{fig1ab}
\caption
{Examples of Dirac spectra. The graph on the left depicts a case of $m>-s$, while
the one on the right is of $m<-s$.}
\end{figure}
%

We first consider one dimensional  potential barrier of constant height
located in positive $x$ region, formally given by
\begin{eqnarray}
\label{e06}
V(x) = v \, \Theta(x),
\quad
S(x) = s \, \Theta(x) ,
\end{eqnarray}
where $ \Theta(x)$ is a Heaviside step function which is one for $x>0$ and zero for $x<0$.
We define ``mass excluded'' energy  $w$ by 
\begin{eqnarray}
\label{e07}
w = \varepsilon-m,
\end{eqnarray}
which we assume to be positive.
The spectra of a Dirac particle inside the potential barrier is composed of two disjoint
continuous spectra separated by a gap.  Some examples of the spectra are shown in FIG. 1.
The scattering wave functions at $x<0$ and $x>0$ are given,
respectively as
\begin{eqnarray}
\label{e08}
\!\!\!\!\!\!\!\!&&
\pmatrix{ \varphi \cr \chi } 
=\pmatrix{ 1 \cr \frac{ik}{m+\varepsilon} } e^{ikx}
- R \pmatrix{ 1 \cr \frac{-ik}{m+\varepsilon} } e^{-ikx} ,
\nonumber \\
\!\!\!\!\!\!\!\!&&
\pmatrix{ \varphi \cr \chi } 
=T \pmatrix{ 1 \cr \frac{ip}{m+\varepsilon+s-v} } e^{ipx} ,
\end{eqnarray}
with the free momentum $k=\sqrt{\varepsilon^2-m^2}$ and the momentum 
$p=\sqrt{(\varepsilon-v)^2-(m+s)^2}$ inside the potential barrier.  
The expression (\ref{e08}) is valid
for the case of continuous spectra for barrier region, $\varepsilon>|m+s|+v$.  For the
case of $|m+s|>\varepsilon>-|m+s|+v$, we need to make replacement,
$p = i \kappa$ with  $\kappa=\sqrt{-(\varepsilon-v)^2+(m+s)^2}$.
The case of $\varepsilon<-|m+s|+v$ corresponds to a particle under the Dirac sea,
for which a  $p \to -p$ is needed in (\ref{e08}), but this case is soon to be  sown irrelevant.
Reflection and transmission rate is given by the squared absolute values of
coefficients $R$ and $T$, respectively, when $p$ is real.  
When $\kappa$ is real,
on the other hand, $T$ is the amplitude of wave function at classically forbidden region.

Expressing the momenta in terms of the energy $w$,
we have
\begin{eqnarray}
\label{e10}
&&
k = \sqrt{w(w+2m)},
\nonumber\\
&&
p = \sqrt{ (w-s-v) (w+2m+s-v) } .
\end{eqnarray}
The matching of wave functions (\ref{e08}) at $x=0$ gives
\begin{eqnarray}
\label{e11}
1-R = T,
\quad
g (1+R) = T
\end{eqnarray}
with
\begin{eqnarray}
\label{e12}
g \! 
=  \!\! \sqrt{ \!\!\frac{ w(w\!+\!2m\!+\!s\!-\!v) }{ (w\!+\!2m)(w\!-\!s\!-\!v) } } Q
\!=\!\!  \frac{1}{i} \sqrt{\!\! \frac{ w(w\!+\!2m\!+\!s\!-\!v) }{ (w\!+\!2m)(s\!+\!v\!-\!w) } } Q.
\end{eqnarray}
Here, $Q$ is the  {\it Giachetti-Sorace factor} given by
\begin{eqnarray}
\label{e12aa}
Q = 1-\Theta(v-s-2m-w) \Theta(s+v-w).
\end{eqnarray}
that represents the exclusion of wave function to the barrier region $x>0$
when the energy $w$ hits the negative energy spectra of Dirac equation with potentials
$s$ and $v$.
It is technically obtained from the proper connection condition
$\varphi(0_-)=0$, $\chi(0_-)=constant$, which is  obtainable as
the $n \to \infty$ limit of $(x-1)^n$ potential, for
which $\varphi(0_-)=0$, $\chi(0_-)=constant$ is found to be the 
correct condition \cite{GS07}. 
Fuller picture of this peculiar boundary condition 
may require the treatment of the problem with proper
field theoretical setup as in \cite{KS04},
where the exclusion factor $Q$ could be understood as a result of many-body Pauli blocking.

The solution of the problem (\ref{e11}) is elementary, and we have
\begin{eqnarray}
\label{e13}
T=\frac{2g}{1+g},
\quad
R = \frac{1-g}{1+g}.
\end{eqnarray}
%
For large enough $w$, that satisfies the condition $w> |m+s| -m+v $, 
the spectra inside the potential region is continuous, and 
we have patial transmission and reflection specified by (\ref{e13}) with (\ref{e12}).  
We naturally have unitarity relation $|R|^2+|T|^2=1$.
As we decrease $w$ down to the threshold energy $w=v-m+|m+s|$, $p$ approaches zero,
and 
$g$ becomes either zero (if $s < -m$ and therefore $w=v-s-2m$) or infinity
(if $s > -m$ and therefore $w=v+s$). They respectively correspond to perfect reflection $R=1$ with
Dirichlet boundary $\varphi(0_-)=0$ or $R=-1$ with Neumann boundary $\chi(0_-)=0$.

Below this threshold, $ |m+s| -m+v > w  > - |m+s| -m+v $ (or $|\varepsilon-v|<|m+s|$,
if it occurs with positive $w$), 
we have exponential wave function with decay constant $\kappa$. 
The full reflection $|R|=1$ with quantum penetration $0 < T < 2$ to the classically forbiden area
is observed. 
Note that there is no problem in having $|T|>1$ in this case, since the unitarity is guaranteed by
decaying wave function $e^{-\kappa x}$.
At the ``Dirac sea'' threshold, $w=v-m-|m+s|$, $\kappa$ approaches zero, and 
$g$ becomes either infinity (if $s < -m$ and therefore $w=v+s$) or zero
 (if $s > -m$ and therefore $w=v-s-2m$). 
They again respectively correspond to perfect reflection $R=-1$ with
Neumann boundary $\chi(0_-)=0$ or $R=1$ with Dirichlet boundary $\varphi(0_-)=0$.

Below the Dirac sea threshold, $w  < - |m+w| -m+v$  we have perfect reflection with
Dirichlet  condition $R=1$ as a result of Giachetti-Sorace factor.
  
When the energy $w$ approaches $0$, we have $\kappa \to 0$, that signifies
the quantum penetration length to classically forbidden region $x>0$ becoming infinite.
However, (\ref{e12})  tells us that we have $g=0$, and thus no penetration amplitude
$T=0$ and, as a result, the perfect ``classical'' reflection $R=1$.

%
\begin{figure}
\center{ 
\includegraphics[width=8.8cm]{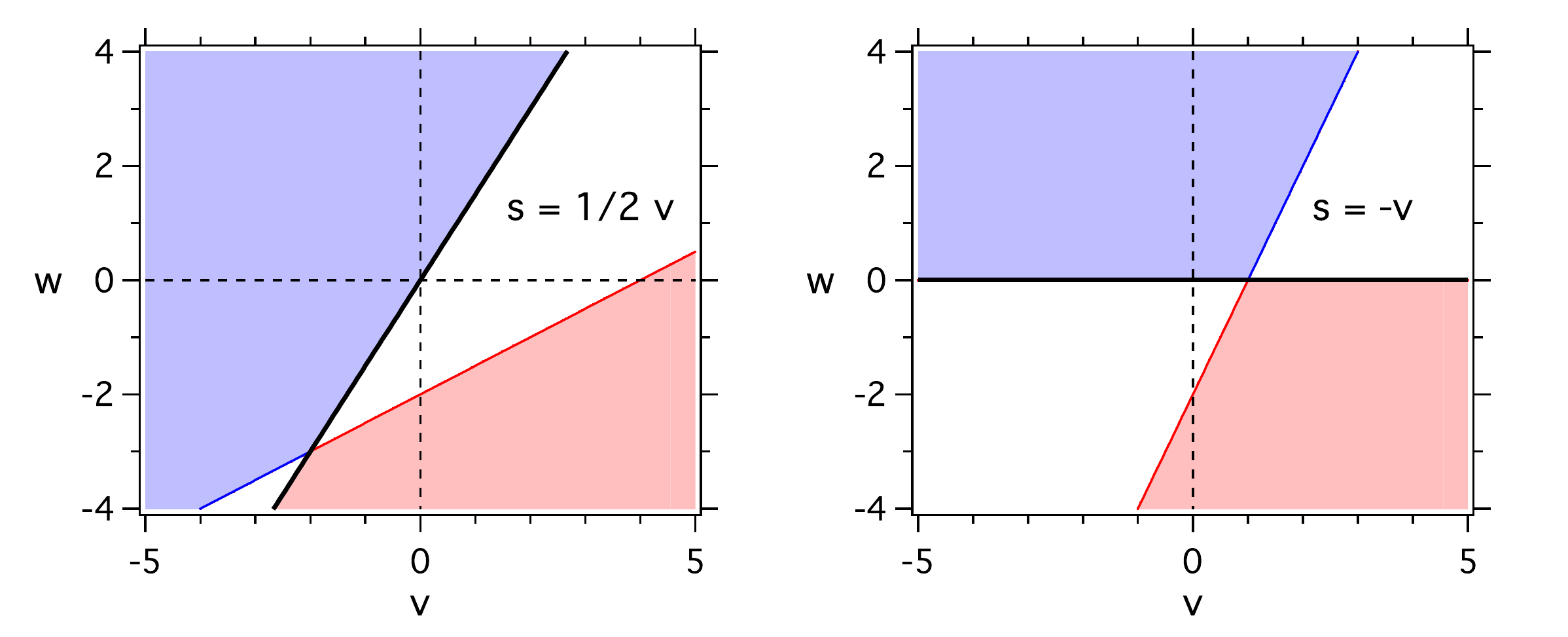}
}
\label{fig2}
\caption
{Examples of Dirac spectra under scalar and vector potentials, $s$ and $v$.
Here, $s$ is chosen to be proportional to $v$.  The blue region represents the 
continuous particle spectra, while the red region is the ``Dirac sea'' spectra,
which is occupied for the vacuum state.  White region is the gap.  Mass is set to be $m=1$.
Among the lines separating spectra, solid line represents $w = v+s$, on which
we have  $g=\infty$ 
and the other line $w=v-s-2m$ on which we have $g=0$.
At the crossing point of these two lines, $g$ takes the value $g=\sqrt{\frac{v+s}{v-s}}$.
}
\end{figure}
%
Above statements are true in generic case, depicted in the left hand graph in
FIG. 2, for example,
but there is an exception to the case when potentials $v$ and $s$ 
are related by $s + v = 0$, depicted in the right hand graph of FIG. 2.
In this case, there is a cancelation in the expression for $g$, (\ref{e12}) and we have 
\begin{eqnarray}
\label{e17}
g =  \sqrt{ \frac{ w\!+2m\!-2v }{ w\!+2m } } 
  =\frac{1}{i}\sqrt{ \frac{ 2v\!-w\!-2m }{ w\!+2m } } .
\end{eqnarray}
which is finite as a results of ``merging'' of
$w=v+s$ line, on which $g=\infty$ holds, and $w=0$ line, on which $g=0$ holds.

This means that for this special case of opposite-sign but equal-magnitude scalar 
and vector potential, $s+v=0$, we have singular infinite-range penetration limit $\kappa\to0$
with finite amplitude $0<|T|<2$ for zero energy barrier reflection $w\to0$.  
We rush to note that this poses no
paradox of any sort, since we still have full reflection $|R|=1$ albeit with some nontrivial
phase for $R$.  
This is a subtle but an exotic exception, nonetheless, whose significance soon becomes
obvious in the following.

%
%


We now consider one dimensional scattering by square well of constant height potential
with spatial extension $L$ (FIG. 3), formally given by
\begin{eqnarray}
\label{e21}
V(x) = v \, \Theta(x)\Theta(L-x) ,
\quad
S(x) = s \, \Theta(x)\Theta(L-x) .
\end{eqnarray}
The scattering wave functions at $x<0$, $0<x<L$ and $L<x$ are given,
respectively as
\begin{eqnarray}
\label{e22}
&&\pmatrix{ \varphi \cr \chi } 
=\pmatrix{ 1 \cr \frac{ik}{m+\varepsilon} } e^{ikx}
- R \pmatrix{ 1 \cr \frac{-ik}{m+\varepsilon} } e^{-ikx} ,
\nonumber \\
&&\pmatrix{ \varphi \cr \chi } 
=A \pmatrix{ 1 \cr \frac{ip}{m+\varepsilon+s-v} } e^{ipx}
- B \pmatrix{ 1 \cr \frac{-ip}{m+\varepsilon+s-v} } e^{-ipx} ,
\nonumber \\
&&\pmatrix{ \varphi \cr \chi } 
=T \pmatrix{ 1 \cr \frac{ik}{m+\varepsilon} } e^{ikx} .
\end{eqnarray}
In a similar manner to the previous case, smooth
connection conditions for both large and small components 
at $x=0$ and $x=L$ gives
\begin{eqnarray}
\label{e24}
&&
1-R = A - B 
\nonumber \\
&&
g (1+R) = A+B
\\ \nonumber
&&
A e^{ipL} - B e^{-ipL} =  T e^{ikL} 
\\ \nonumber 
&&
A e^{ipL} + B e^{-ipL} =  g T e^{ikL} .
\end{eqnarray}
%
%
\begin{figure}
\center{ \includegraphics[width=4cm]{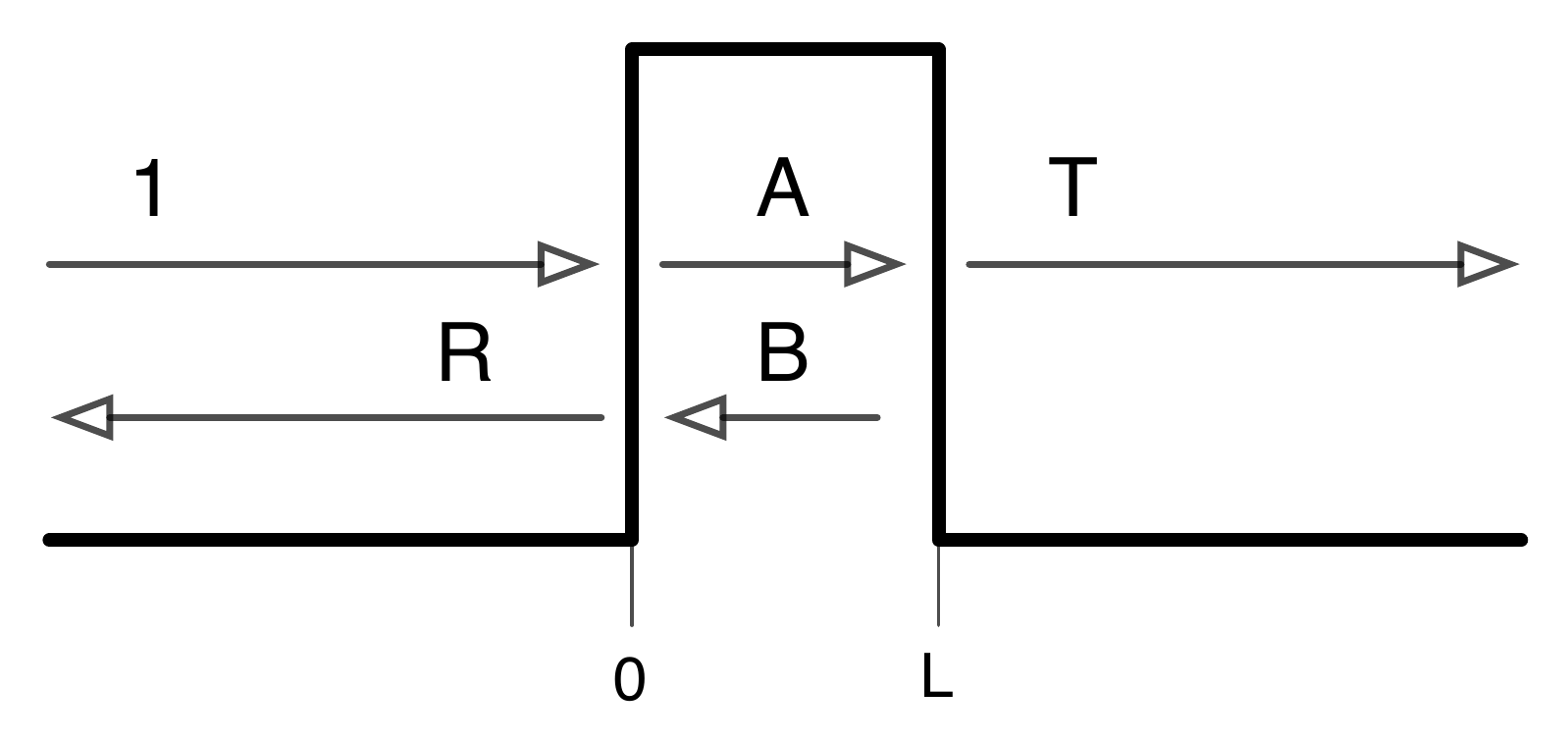} }
\label{fig3}
\caption
{Schematic representations of relativistic potential barrier scattering}
\end{figure}
%
%

Elementary calculation yields the following expressions for transmission and reflection
amplitudes;
\begin{eqnarray}
\label{e26}
&&
T =  \frac {e^{-ikL}} {\cos pL -\frac{i}{2} (1/g+g) \sin pL } ,
\nonumber \\&&
R =  \frac {-\frac{i}{2} (1/g-g) \sin p L} {\cos pL -\frac{i}{2}  (1/g+g)\sin pL } .
\end{eqnarray}
This expression is literally valid for the energy $w>|m+s|+v$. 
For the energy $|m+s|+v>w>-|m+s|+v$, 
we have to make replacement $p = i \kappa$ as before, which will result
in the replacements $\cos pL \to  \cosh \kappa L$ 
and $\sin pL \to i \sinh \kappa L$ in (\ref{e26}).
%
%
This expression is reduced to  $T=0$, $R=1$
%
%
for the energy $w<-|m+s|+v$ with which 
we hit the Dirac sea spectra inside the potential barrier,
where we have $Q=0$, thus $g=0$.


We look at the low energy limit of the scattering matrix $T$.  Generically, for
the case of $s \ne -v$, we have the quantity $g$ that approaches to zero
as we take $w \to 0$ limit,  
causing the divergence of $1/g$, which guarantees the
perfect reflection
\begin{eqnarray}
\label{e31}
T\to 0, \ \  R \to 1
 \ \ {\rm as}\ \ w \to 0 
\qquad (s+v \ne 0).
\end{eqnarray}
This simply is a exact expression of the intuitive statement that 
a generic obstacle works as a reflecting block for low energy
projectile, or in other word, if we hit any barrier 
too slowly, we are bound to get reflected all the time.

However, for the special case of  $s + v = 0$,  $g$ takes the form (\ref{e17}) after
cancelation of $w$ in both denominator and enumerator of (\ref{e12}), and we have 
$g={\rm finite}$ and $\kappa \to 0$ (or $p \to 0$) as we take $w \to 0$ limit.
We therefore obtain, from  (\ref{e26}), a peculiar limit
\begin{eqnarray}
\label{e32}
T \to 1, \ \ R \to 0  \ \ {\rm as}\ \ w \to 0 \qquad (s+v = 0),  
\end{eqnarray}
which signifies an anomalous {\it full transmission at zero energy}.  
This is particularly intriguing for 
the case of decaying wave in the gap region, in which $\kappa$ is real, 
where decaying length $1/\kappa$ becomes
infinity at $w \to 0$ limit. 

The situation is immediately understood by inspecting the illustrations in FIG. 4,
Here, the graph in the left depicts a generic case that has normal perfect reflection
at $w\to0$ limit, while the graph in the right shows the anomalous 
zero energy transparency.
The reason
behind this transparency lies in the enhanced long range tunneling
inside the barrier, which occurs because, at $w \to 0$,
the energy approaches to the threshold of negative continuous spectra
that exists right below $w = 0$ for $s+v=0$ potentials.  
The presence of
Dirac sea not only induces the perfect reflection for $w>0$ with
$v>|s|$, for example,
it also affects the decaying length and induces the anomalous
tunneling, and transmission at $w \to 0$ limit for the case of $s+v=0$ potentials.
%
\begin{figure}
\center{ \includegraphics[width=8.8cm]{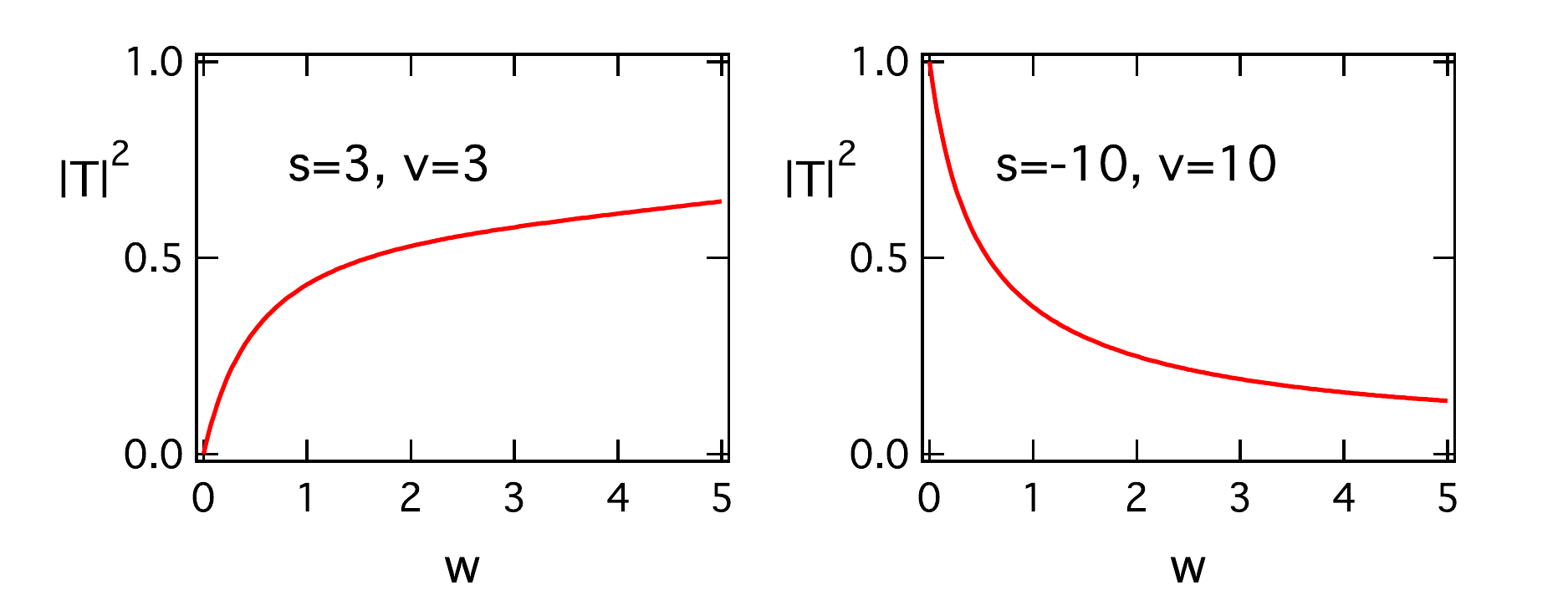} }
\label{fig4}
\caption
{Transmission rate as function of kinetic energy.  The graph on the left is for normal
case, and the one on the right is anomalous case, $s=-v$.}
\end{figure}
%

Let us now consider the $L \to 0$ limit of relativistic scattering. 
Straightforward limit will, of course, lead to disappearance of the barrier, $T \to 1$.  
Limit $L \to 0$ with constant volume integrals,
\begin{eqnarray}
\label{e37}
v= \frac{\bar v}{L}, \quad s = \frac{\bar s}{L}, \qquad (L \to 0),
\end{eqnarray}
on the other hand, leads us to
\begin{eqnarray}
\label{e38}
&&\!\!\!\!\!\!\!\!\!\!
T =  
\frac {1} 
  {\cos \beta +\frac{i}{2\beta} \sin \beta 
                \left[ ({\bar v}\!+{\bar s})/K\!+({\bar v}\!-{\bar s})K \right] 
  } \Theta({\bar v}+|{\bar s}|),
\nonumber \\
&&\!\!\!\!\!\!\!\!\!\!
R =  
\frac {\frac{i}{2\beta} \sin \beta 
                \left[ ({\bar v}\!+ {\bar s})/K\!-({\bar v}\!- {\bar s})K \right]} 
  {\cos \beta +\frac{i}{2\beta} \sin \beta 
                \left[ ({\bar v}\!+ {\bar s})/K\!+({\bar v}\!- {\bar s})K \right] 
  } \Theta(\!-{\bar v}\!-\!|{\bar s}|)
\nonumber \\
&&\!\!\!\!\qquad\qquad\qquad
+\Theta({\bar v}+|{\bar s}|),
\end{eqnarray}
with $\beta = \sqrt{{\bar v}^2-{\bar s}^2}$$=i \sqrt{{\bar s}^2-{\bar v}^2}$ 
and $K=\sqrt{\frac{w}{w+2m}}$.

We can interpret this result in terms of relativistic {\it point interactions} specified by
the boundary condition which is a most general time-reversal symmetric one \cite{AG88}
\begin{eqnarray}
\label{e43}
\pmatrix{ \varphi(0_+) \cr \chi(0_+) }
=
\pmatrix{ \alpha & u_- \cr u_+& \alpha }
\pmatrix{ \varphi(0_-) \cr \chi(0_-) }
\end{eqnarray}
with $\alpha^2-u_+ u_- = 1$.  The scattering off the point interaction (\ref{e43}) is  given
\begin{eqnarray}
\label{e44}
T \!=\! \frac{1} {\alpha\!+\!\frac{i}{2}
[ u_+ /K \! - u_- K ] } ,
\ \!
R \!=\! \frac{ \frac{i}{2}
[ u_+ /K \! + u_-K ] } 
{\alpha\!+\!\frac{i}{2}
[ u_+/K \! - u_- K ] } ,
\end{eqnarray}
which allows the identifications 
\begin{eqnarray}
\label{e45}
u_+ = ({\bar s}+{\bar v}) \frac{\sin\beta}{\beta} ,
\quad
u_- = ({\bar s}-{\bar v}) \frac{\sin\beta}{\beta} .
\end{eqnarray}
The special cases ${\bar s} = {\bar v}$ and ${\bar s} = -{\bar v}$ can be 
considered as the limiting cases of (\ref{e38}), and we have,  for ${\bar s} = {\bar v}$
\begin{eqnarray}
\label{e41}
T = \frac{1} {1+i {\bar v}/K } ,
\quad
R =  \frac{i {\bar v} /K} {1+i {\bar v} /K } ,
\end{eqnarray}
while, for ${\bar s} = -{\bar v}$, we have
\begin{eqnarray}
\label{e42}
T = \frac{1} {1+i {\bar v} K } ,
\quad
R =  \frac{-i {\bar v} K} {1+i {\bar v} K } .
\end{eqnarray}
If we take the {\it non-relativistic limit in kinematics}, $K \to k/(2m)$, these
two cases are exactly identical to the scattering form {\it delta} and {\it delta-prime} point 
interactions \cite{SE86}, which represent high-pass and low-pass wave filters, respectively.
Note that constructing non-standard point interactions, that results in (\ref{e42}), 
within non-relativistic framework involves highly singular procedures \cite{AE94,CS98}.

%

Finally, we ask a question whether we can construct an analogue of 
the phenomena we have found in the framework of
non-relativistic Schr{\"o}dinger equation. 
We rewrite the Dirac equation (\ref{e05}) by eliminating small component $\chi$ in the form
\begin{eqnarray}
\label{e51}
-\frac{d}{dx}\frac{1}{2m^*}\frac{d}{dx}\varphi +U \varphi = w \varphi ,
\end{eqnarray}
with effective mass $m^*$ and potential $U$ defined by
\begin{eqnarray}
\label{e52}
m^* =  m + \frac{w}{2} + \frac{S - V}{2},
\quad
U = S + V .
\end{eqnarray}
Assuming the conditions $w << m$ and $|S-V| << m$, we obtain  Schr{\"o}dinger equation
with effective potential which is given by the sum of vector and scalar potentials. This
is nothing but the 
true non-relativistic limit.
However, we obtain non-standard low-energy limit 
by assuming the non-relativistic kinematics $w << m$ in conjunction
with strong relativistic potentials  $|S-V| \sim m$.   
Specifically, we can reproduce anomalous transmission 
from Schr{\"o}dinger equation (\ref{e51}) by setting $S=-V$ 
which results in $m^* \approx m-V$ and $U=0$.  This means that we can construct a purely
non-relativistic model of anomalous scattering and delta-prime point interaction
with {\it just effective mass and no potential}.  Readers are warned, however, that this non-relativistic
analogue scheme works only to an extent:  
When $S-V$ is negative in sign and so large, we obtain negative
value for the effective mass $m^*$. 
For this {\it bona fide} relativistic dynamics, non-relativistic analogue (\ref{e51}) 
does not make sense, and therefore does not exist.

%
We have shown that a set of relativistic potentials having the property of 
same strength but opposite signs for scalar and vector components displays
anomalous full tunneling and transparency at zero energy, while barrier starts functioning 
at higher energy.  The short range limit of this phenomenon leads to a smooth
relativistic realization of an exotic point interaction, delta-prime, that 
conventionally requires singular and esoteric constructions
within non-relativistic dynamics. 
It has been pointed out \cite{GI97} that in three dimensions, the ``$S=-V$'' relativistic potentials 
have an esoteric property called {\it pseudospin symmetry} \cite{AH69} that
has been found to play important role in the degeneracy structure of nuclear levels.  
Current work shows that there is yet another aspect
to this pseudospin symmetric limit of relativistic potentials, 
which is revealed only in one dimensional systems.
%

\acknowledgments
We would like to thank Prof. Riccardo Giachetti for enlightening discussions.
We also wish to thank Prof. Hiroshi Frusawa and Prof. Azhar Iqbal for useful comments.
This work has been partially supported by 
the Grant-in-Aid for Scientific Research of  Ministry of Education, 
Culture, Sports, Science and Technology, Japan
under the Grant number 18540384.
 




%
%
%

\end{document}